\documentclass{optica-article}

\journal{opticajournal} 

\articletype{Research Article}

\usepackage{lineno}

\usepackage{siunitx}
\usepackage{caption}
\usepackage{subcaption}
\usepackage[]{hyperref}
\hypersetup{
	colorlinks=true,
	linkcolor=blue,
	citecolor=blue,
	urlcolor=blue
}
\usepackage{ulem}
\usepackage{braket}
\usepackage{graphicx,import}
\usepackage{tabulary}
\usepackage{tikz}
\usepackage{booktabs}
\usepackage{titlesec}

\newsavebox{\imagebox}

\begin{document}
	
	\title{Towards Fully Passive Time-Bin Quantum Key Distribution over Moving Free-Space Channels}
	
	\author{Ramy Tannous\authormark{1,2,*,$\dagger$}, Wilson Wu\authormark{1,2,3}, St\'ephane Vinet\authormark{1,2}, Chithrabhanu Perumangatt\authormark{4},\\ Dogan Sinar\authormark{1,2}, Alexander Ling\authormark{4,5}, Thomas Jennewein\authormark{1,2,3}}
	
	\address{\authormark{1}Institute for Quantum Computing, University of Waterloo, 200 University Ave W, Waterloo, Ontario, N2L 3G1, Canada\\
		\authormark{2}Department of Physics \& Astronomy, University of Waterloo, 200 University Ave W, Waterloo, Ontario, N2L 3G1, Canada\\
		\authormark{3}Department of Physics, Simon Fraser University, 8888 University Dr W, Burnaby, BC V5A 1S6, Canada\\
		\authormark{4}Centre for Quantum Technologies, 3 Science Drive 2, National University of Singapore 117543 Singapore\\
		\authormark{5}Department of Physics, Faculty of Science, National University of Singapore, 2 Science Drive 3, 117551 Singapore\\
		\authormark{$\dagger$}Current Address: National Research Council of Canada, 100 Sussex Drive, Ottawa, Ontario, K1A 0R6, Canada}
	
	\email{\authormark{*}ramy.tannous@uwaterloo.ca}
	
\textcopyright{} 2025 Optica Publishing Group under the terms of the \href{https://opg.optica.org/content/library/portal/item/license_v2#VOR-OA}{Optica Open Access Publishing Agreement}. \\DOI:~\url{https://doi.org/10.1364/OE.558610}

	\begin{abstract}
		Encoding quantum information in photonic time-bin states is typically considered impractical for moving free-space quantum communication due to the difficulties with phase stabilization of distant quantum time-bin interferometers and turbulence of free-space channels. We demonstrate a novel approach using reference frame independent time-bin quantum key distribution that completely avoids the need for active relative phase stabilization while simultaneously overcoming a highly multi-mode channel without any active mode filtering. This scheme enables passive, self-compensating time-bin quantum communication without any mode filtering, mode sorting, adaptive optics, active basis selection, or active phase alignment. We realize a proof-of-concept demonstration using hybrid polarization and time-bin entangled photons that demonstrates a sustained asymptotic secure key rate greater than \SI{0.07}{bits\per\text{coincidence}} over a \SI{15}{\meter} multi-mode fiber optical channel and showing entanglement correlations over a moving \SI{38.5}{\decibel} loss free-space channel, including system losses. The scheme simplifies the use of time-bin encoding and can be readily applied over various spatially multi-mode and fluctuating channels involving rapidly moving platforms, including airborne and satellite systems.
	\end{abstract}
	
	\section{Introduction}
	Time-bin encoding has emerged as a favored technique in fiber-based quantum networks~\cite{brendel_pulsed_1999, tittel_quantum_2000, thew_experimental_2002} and in photonic chips~\cite{sibson2017chip}. In typical time-bin systems, photonic state encoding and decoding is usually accomplished with single-mode interferometers and typically employ active phase compensation~\cite{dixon_gigahertz_2008,xavier2011stable,lucamarini2013efficient,Toliver:15,fitzke2022scalable,morales2023optical}. However, when dealing with time varying multi-mode channels (e.g. moving free-space channels), a conventional time-bin receiver necessitates the coupling to single-mode fibers, which can have high losses unless mitigating adaptive optics are employed. Systems that employ active stabilization typically rely on secondary signals that are subject to signal fluctuations and dropouts due to pointing accuracy and turbulence that are amplified over long range moving free-space channels~\cite{ricklin2006atmospheric,dix2021ultra}. Such signal dropouts and fluctuations can be particularly detrimental for the phase stabilization of the sender and receiver interferometers. In this work, we present a novel time-bin quantum communication scheme that is suitable for moving free-space channels and is fully passive, i.e. operates without the use of active basis selection, mode filtering, adaptive optics, or interferometer stabilization. Our approach maintains entanglement correlations in the presence of spatial multi-mode distortions and relative phase drifts between the sender and receiver time-bin interferometers. We demonstrate that over a lower loss channel (multi-mode fiber), our approach achieves a constant quantum bit error ratio (QBER) below $3.5\%$ and $5.5\%$ in the key distillation and phase estimation bases, respectively, while maintaining an asymptotic key rate of greater than \SI{0.07}{bits\per\text{coincidence}} despite intentional phase changes between the sender and receiver interferometer. While over a high loss channel moving free-space channel, our approach is able to maintain constant entanglement correlations.

	To operate over a moving free-space channel, we utilize passive field widened interferometers that have been demonstrated to support quantum key distribution (QKD) over a highly multi-mode channel~\cite{jin2019genuine,jin_demonstration_2018}. The interferometers don't suffer from mode-filtering losses, and are less complex than employing adaptive optics solutions. This is especially beneficial for time-bin channels with moving platforms as between aircraft~\cite{pugh_airborne_2017}, drones, and satellites~\cite{liao_satellite--ground_2017,podmore_qkd_2021} where high pointing errors and mode distortions are present. Our scheme uses the 6-state 4-state time-bin reference frame independent quantum key distribution (RFI-QKD) protocol~\cite{tannous2019demonstration}, which avoids the requirement for active phase locking of the transmitter and receiver interferometers. While the relative phase between the two interferometers drifts, the RFI-QKD protocol maintains a secret key rate provided the phase drift is slow enough for the estimation of channel parameters. Recently, several time-bin RFI-QKD demonstrations have been performed over static free-space and optical fiber channels~\cite{chen_field_2020,feng_four-state_2021,tang_free-running_2022}. For free-space channels, RFI-QKD protocols are advantageous since removing the active phase alignment eliminates the difficulties associated with the random intensity fluctuations a reference signal would suffer over a turbulent free-space channel. Although power normalization of the reference signal is possible, it typically requires advanced techniques and additional resources ~\cite{guiomar2022coherentModelling,vetelino2007fade,abtahi2006suppressionSOA,khalighi2014survey}. Furthermore, this scheme could also avoid the need to compensate for Doppler shift induced phase drift in a satellite link~\cite{space_super_zeitler2016super,vallone2016interference,chapman2020time}. Thus, our fully passive scheme which employs the combination of RFI-QKD and a field widened interferometer can be used to achieve the distribution of a secret key over a multi-mode channel, as well as resolve a sufficiently slow relative phase drift, and a channel phase drift without active compensation or filtering.

	\section{Experiment}
	For our demonstration we use a polarization entangled photon source (EPS) where one photon from each photon pair is converted from polarization to time-bin~\cite{ma2009experimental}. After the conversion, the state shared between Alice and Bob is
	\begin{equation}
		\label{eq:Estate}
		\ket{\psi_{\text{AB}}}=\frac{1}{\sqrt{2}}\left(\ket{H\mathcal{E}}+e^{i\phi_{\text{AB}}}\ket{V\mathcal{L}}\right),
	\end{equation}
	where $\phi_{AB}=\phi_{HV_A}+\phi_{\mathcal{E}\mathcal{L}_B}$ is the total relative phase between Alice's and Bob's part of the state; $H$, $V$ are the horizontal and vertical polarizations respectively; and $\mathcal{E}$, $\mathcal{L}$ are the early and late time-bins respectively. The particular benefit of this hybrid entangled state is the relative ease of producing a 6-state analyzer for polarization encoded photons, while a 4-state analyzer for time-bin photons is directly implemented by an unbalanced interferometer. Alice performs polarization measurements in the $Z_{A}\text{, } X_{A}\text{, and }Y_{A}$ bases, while Bob performs time-bin measurements in the $Z_{B}\text{ and } X_{B}$ bases. In our implementation, the fixed computational basis is the horizontal-vertical polarization basis ($Z_{A}$) for Alice and the early-late ($Z_{B}$) basis for Bob. Although we use an entanglement based approach, this scheme could easily be adapted for a prepare-and-measure type scheme where the sender produces 6 states while the receiver measures 4 states or vice-versa.
	
	As shown using polarization encoding in~\cite{tannous2019demonstration}, a secret key can be generated despite the drifts of $\phi_{AB}$. This is provided that $\ket{\psi_{\text{AB}}}$ maintains strong correlations in every basis. In particular for our experiment, the strength of the all the superposition correlations is inferred by observing the correlations between the two polarization superposition bases and the time-bin superposition basis. The strength of these correlations can be represented by the $C$-parameter given by
	
	\begin{equation}
		\label{eq:C-param}
		C=\sqrt{\braket{X_{A}\otimes X_{B}}^2+\braket{Y_{A}\otimes X_{B}}^2},
	\end{equation}
	where $0\leq C\leq 1$ allows the users to estimate the channel quality with $C=1$ indicating a maximally entangled state or perfect channel~\cite{tannous2019demonstration}.
	
	The $C$-parameter is phase independent despite any drifts or fluctuations in $\phi_{AB}$, provided the changes do not occur too rapidly~\cite{wang2016valid,yoon_experimental_2019}. The degree to which the phase can change is a function of the photon detection rate and timing resolution of the system.
	
	
	\begin{figure}[htbp]
		\centering
		
		\includegraphics[width=0.99\textwidth]{experimental_setup.pdf}
		
		\caption{Experimental setup of the time-bin RFI-QKD proof-of-concept demonstration. An entangled photon source (EPS) creates polarization entangled photon pairs at \SI{785}{\nano\meter} (signal) and \SI{842}{\nano\meter} (idler) via type-0 spontaneous parametric downconversion. The idler photon is measured at a local 6-state polarization analyzer, while the signal photon is converted to time-bin encoding by the polarization to time-bin converter (PTC) and sent across the quantum channel: a \SI{15}{\meter} multi-mode fiber (MMF) channel or a moving free-space channel. The time-bin analyzer (TA) performs a 4-state time-bin measurement. Both the PTC and TA interferometers have a path separation of \SI{2}{ns}. A piezoelectric actuator is inserted in the TA short path to induce phase modulations. The input and output of the TA are MMF. HWP: half-waveplate, QWP: quarter-waveplate, L: lens, POL: polarizer, PMF: polarization maintaining fiber, PBS: polarizing beam splitter, BS: 50:50 beam splitter.}
		\label{fig:setup}
	\end{figure}

	\begin{figure}[!hbtp]
		\centering
		\begin{subfigure}[c]{0.495\textwidth}
			\centering
			\includegraphics[width=\columnwidth]{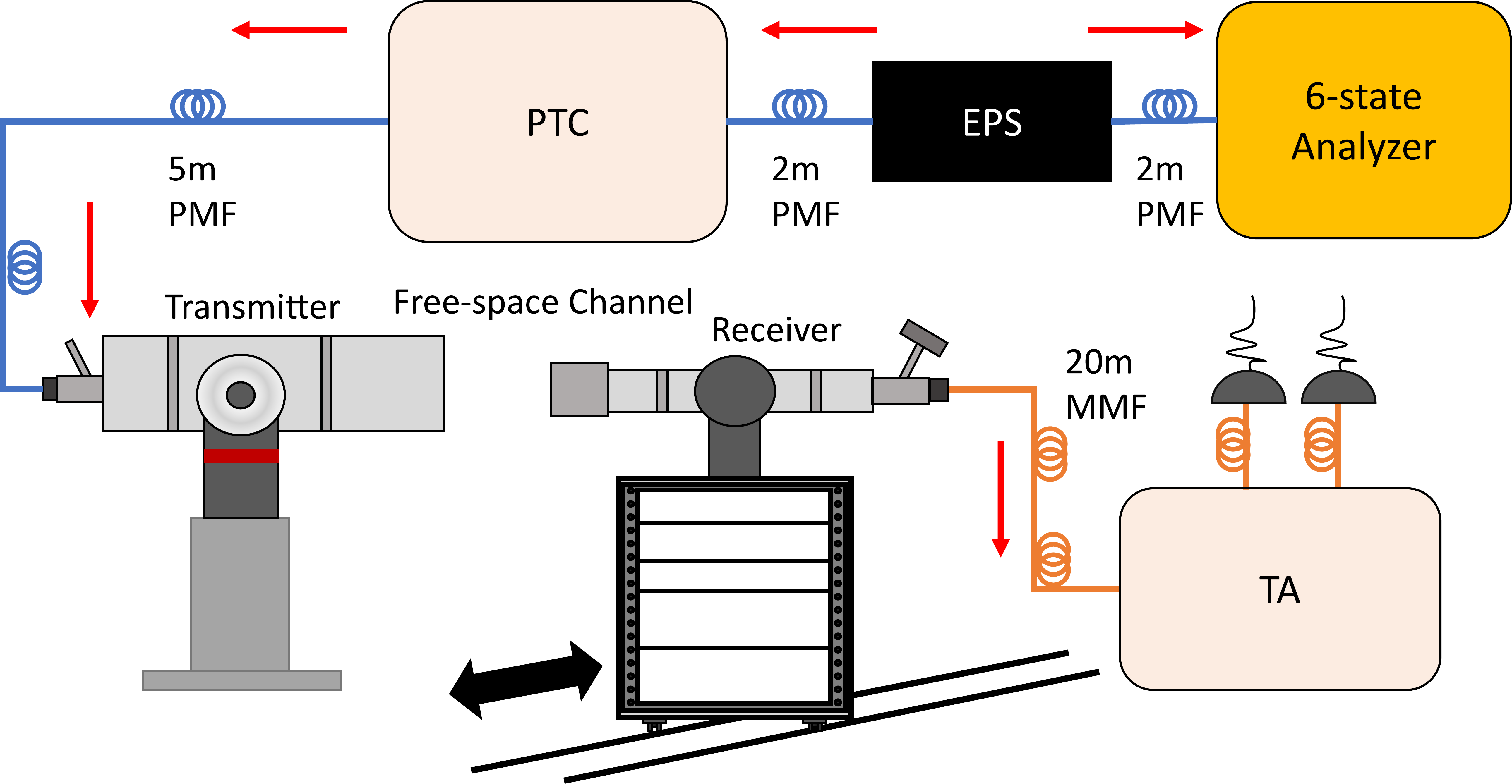}
			
			\caption{}
			\label{fig:block_receiver}
		\end{subfigure}
		\begin{subfigure}[c]{0.495\textwidth}
			\centering
			\includegraphics[height=4cm]{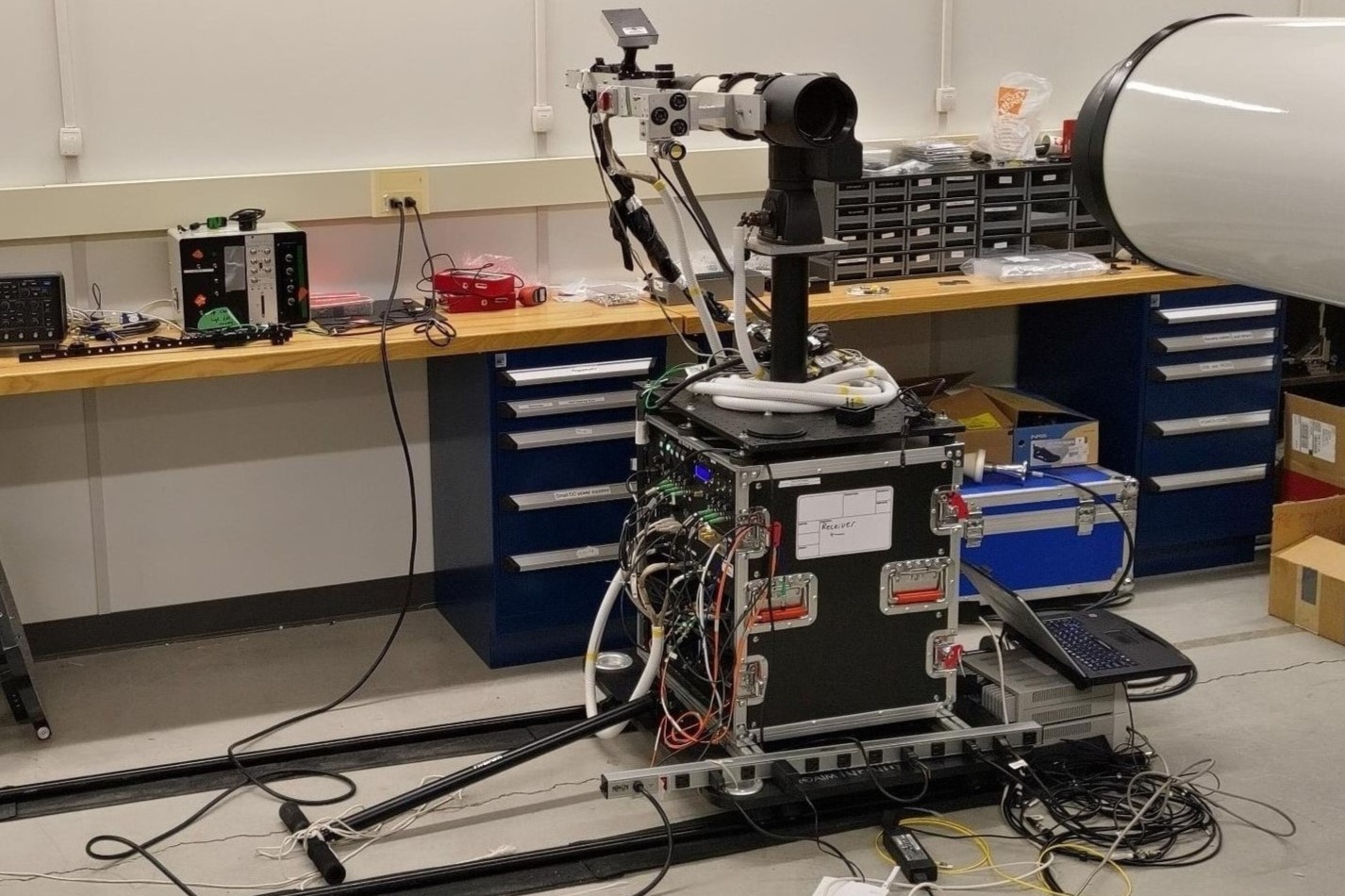}
			\caption{}
			\label{fig:real_receiver}
		\end{subfigure}
		\caption{(\protect\NoHyper\subref{fig:block_receiver}\protect\endNoHyper) Moving free-space channel experimental setup. The rail system allows a user to move the receiver system along a \SI{6.5}{\meter} track. EPS: Entangled photon source. PTC: Polarization to time-bin converter. TA: Time-bin analyzer. (\protect\NoHyper\subref{fig:real_receiver}\protect\endNoHyper) Moving receiver setup on the rail system.}
		\label{fig:tb_moving_setup}
	\end{figure}
	
	Fig.~\ref{fig:setup} shows a sketch of the experimental setup used for the time-bin RFI-QKD proof-of-concept demonstration, while Fig.~\ref{fig:tb_moving_setup} shows the setup when being used with a moving free-space channel along a \SI{6.5}{\meter} track~\cite{tannous2023advancing}. The entangled photon source~\cite{lohrmann2020broadband} creates polarization entangled photon pairs in the state $\ket{\psi_{AB}}=\frac{1}{\sqrt{2}}\left(\ket{H_A H_B}+e^{i\phi_{AB^*}}\ket{V_A V_B}\right)$, where $\phi_{AB^*}$ is a random phase introduced by the EPS. A \SI{406}{\nano\meter} continuous-wave diode laser is used to pump a periodically-poled potassium titanyl phosphate (PPKTP, length: \SI{10}{\milli\meter}, poling period: \SI{3.425}{\micro\meter}) bulk crystal to create signal (\SI{785\pm5}{\nano\meter}) and idler (\SI{842\pm5}{\nano\meter}) photon pairs. The EPS is built in a linear Mach Zehnder type configuration as shown in Ref.~\cite{lohrmann2020broadband}. The signal and idler are separated by a dichroic filter and coupled into polarization maintain fiber (PMF, Thorlabs PM780-HP). The idler photon is measured using a 6-state polarization measurement~\cite{tannous2019demonstration}, while the signal photon is converted to time-bin encoding, by the polarization to time-bin converter (PTC), before being sent across the multi-mode quantum channel. The PTC is an unbalanced Michelson interferometer shown in Fig.~\ref{fig:setup}. The PTC is designed such that the horizontal polarized light takes the short path $(H\rightarrow \mathcal{E})$ and vertically polarized light takes the long path, $(V\rightarrow \mathcal{L})$ which creates the entangled state in (\ref{eq:Estate}). Thus any photon in a polarization superposition would create a superposition state of both interferometer paths in the PTC. The polarization information of the photons is removed at the output of the PTC by placing a polarizer at \ang{45} with respect to the horizontal and vertical polarizations. The time-bins separation given by the path difference within the PTC is approximately \SI{2}{\nano\second}. At the output of the PTC, the photon is coupled to a PMF with the polarization aligned to one of the axes of the fiber. This PMF is then connected to the multi-mode fiber channel or the moving free-space transmitter.
	
	The multi-mode fiber (MMF) channel is a graded-index fiber (Corning InfiniCor 300 62.5/125) which has a core size of \SI{62.5}{\micro\meter} and a numerical aperture of $0.275$. The estimated maximum modal dispersion for this fiber is \SI{0.353}{\nano\second\per\kilo\meter} which amounts to a negligible \SI{5.2}{\pico\second} over the \SI{15}{\meter} length. Despite the relatively short link distance, extensions to farther distances (several kilometers) are possible as permitted by both the modal dispersion and the fiber losses. Furthermore, the use of a multi-mode fiber as the channel is not essential to our demonstration and is used to emulate a spatially multi-mode channel with losses that are tolerable by the current setup. Furthermore, the feasibility of the RFI-QKD time-bin protocol with an MMF eliminates the need for full adaptive optics by use a MMF as the collection fiber for a moving free-space receiver.
	
	The moving free-space channel consists of a transmitter and receiver system, each equipped with a tracking system. This tracking system is composed of two subsystems: a coarse pointing (CP) system and a fine pointing (FP) system which ensure the transmitter and receiver telescopes remain aligned~\cite{pugh_airborne_2017}. Both pointing systems use PID loops that monitors the position and speed of the spot of a \SI{850}{\nano\meter} beacon laser with a camera. The FP system uses a steering mirror that is able to recover up to \SI{0.5}{\degree} in pointing errors and has an accuracy of \SI{50}{\micro\radian}~\cite{pugh_airborne_2017}. The FP unit contains a \SI{50}{\micro\meter} pinhole is used as a spatial filter to limit the amount of stray light that is collected by the system. After passing the spatial filter, the signal passes through a 50:50 beamsplitter that is present due to a two-basis, four output, polarization analyzer for which the receiver system is designed. This polarization analyzer adds additional losses to the channel as only one of four outputs is used to collect the time-bin signals. The quantum signal is then coupled into multi-mode fibers that are at the output of the FP unit. The receiver telescope and pointing system is placed on a trolley and rail system that is moved manually and therefore emulates a moving receiver. Further details on the transmitter and receiver setups can be found in ref~\cite{tannous2023advancing}, ref~\cite{pugh_airborne_2017}, and the supplemental material.
	
	A field widened multi-mode time-bin analyzer (TA) is used to measure the time-bin encoded photons after passing through the quantum channel~\cite{jin2019genuine}. The optical layout of the TA is shown in Fig.~\ref{fig:setup}. The TA uses a $4f$ imaging system in each arm. The relative phase of the PTC and TA are allowed to drift with ambient noise. In some photon exchanges, a piezoelectric actuator is used to deliberately induce phase modulation of the relative path difference between the two paths of the TA, demonstrating the robustness of the protocol (Fig.~\ref{fig: results2}). The outputs of the TA are coupled into multi-mode fibers and detected by silicon single photon avalanche diodes (Excelitas SPCM-AQRH). In the moving free-space channel demonstration, the TA is connected to the end of the receiver telescope using the \SI{15}{\meter} MMF. A time tagging unit (UQDevices Logic-16) is used to record the detection events and coincidence analysis is done to calculate the correlations in the various bases. 
	
	For our proof-of-concept demonstration the measured number of coincidences is approximately \SI{9000}{\text{coincidence}\per\second} for the MMF channel and \SI{150}{\text{coincidence}\per\second} for the moving free-space channel. The EPS has an intrinsic pair production rate of the order of \SI{10e6}{\text{pairs}\per\second}. The heralding efficiency of the source is measured to be $\approx20\%$, including the losses due to fiber coupling into the PMF. Furthermore, the losses contributed by the PTC and TA, and the detector efficiencies also reduce the overall number of coincidences measured. The optical transmission of the PTC, TA, quantum channel, and $6$-state analyzer were measured using an external cavity diode laser (Toptica DL pro) and power meter (Thorlabs PM100). The PTC transmission is measured to be $35\pm1\%$ including fiber coupling into the output PMF, but it has an intrinsic maximum transmission of $50\%$ due to the polarizer that eliminates the polarization signature of each time-bin. Therefore the adjusted optical transmission of the PTC is $70\pm2\%$. The \SI{15}{\meter} multi-mode fiber channel has a transmission of $86\pm2\%$ for \SI{785}{\nano\meter}, while the moving free-space channel alone has an transmission of $0.5\pm0.1\%$. The losses of the moving free-space channel are a combination of the intrinsic losses of the telescopes, the losses due to the finite collection size of the receiver telescope, and the unavoidable intrinsic losses of the polarization analyzer.
	The TA has an optical transmission of $67\pm2\%$, including the coupling losses of the collection fibers. The $6$-state polarization analyzer at Alice has an average transmission of $49\pm2\%$ across all six paths. Experimentally, the total combined efficiencies, including single photon detector efficiency and losses due to spectral filtering, are $9.3\pm0.1\%$ for the idler photons measured by Alice, $1.1\pm0.1\%$ for the signal photons measured by Bob over the MMF channel and $0.014\pm0.001\%$ for the moving free-space channel. Table.~\ref{tab:efficiencies} summaries the efficiencies and their corresponding loss of the experimental subsystems used.

	\begin{table}[]
		\centering
		\caption{Collection efficiency and single photon transmission and loss of various experimental subsystems. The total channel values include detector efficiencies.}
		\label{tab:efficiencies}
		\begin{tabular}{@{}l|c|c@{}}
			\toprule
			Subsystem & Efficiency [\%] & Loss [dB] \\ \midrule
			Entangled photon source &   11-20     &         9.6-7      \\
			Polarization-time bin converter &       35      &     4.56     \\
			15m Multi-mode fiber &  86     &      0.655          \\
			Free-space channel &  0.5    &       23          \\
			Time bin analyzer &     67     &         1.74    \\
			$6$-state polarization analyzer  &    49      &     3.1       \\ \hline
			Total Multi-mode channel  &    1.1      &     19.6        \\ 
			Total Free-space channel  &    0.014      &      38.5       \\ \bottomrule
		\end{tabular}
	\end{table}

	\section{Results}
	
\begin{figure}[htbp]
		\centering
		\begin{subfigure}[t!]{0.495\textwidth}
			\centering
			\includegraphics[width=\textwidth]{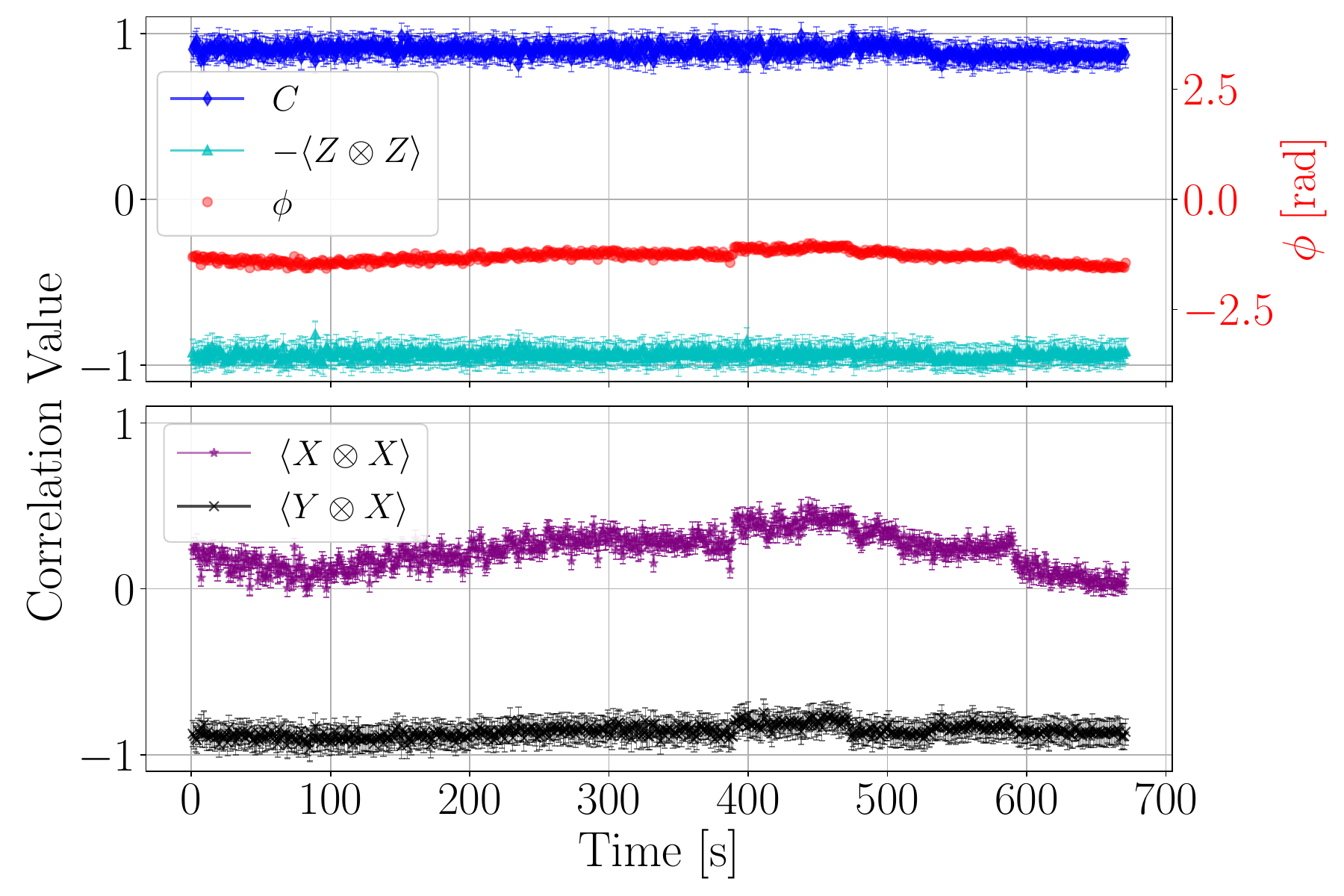} 
			\caption{}
			\label{fig:ex1}
		\end{subfigure}
		\begin{subfigure}[t!]{0.495\textwidth}
			\centering
			\includegraphics[width=\textwidth]{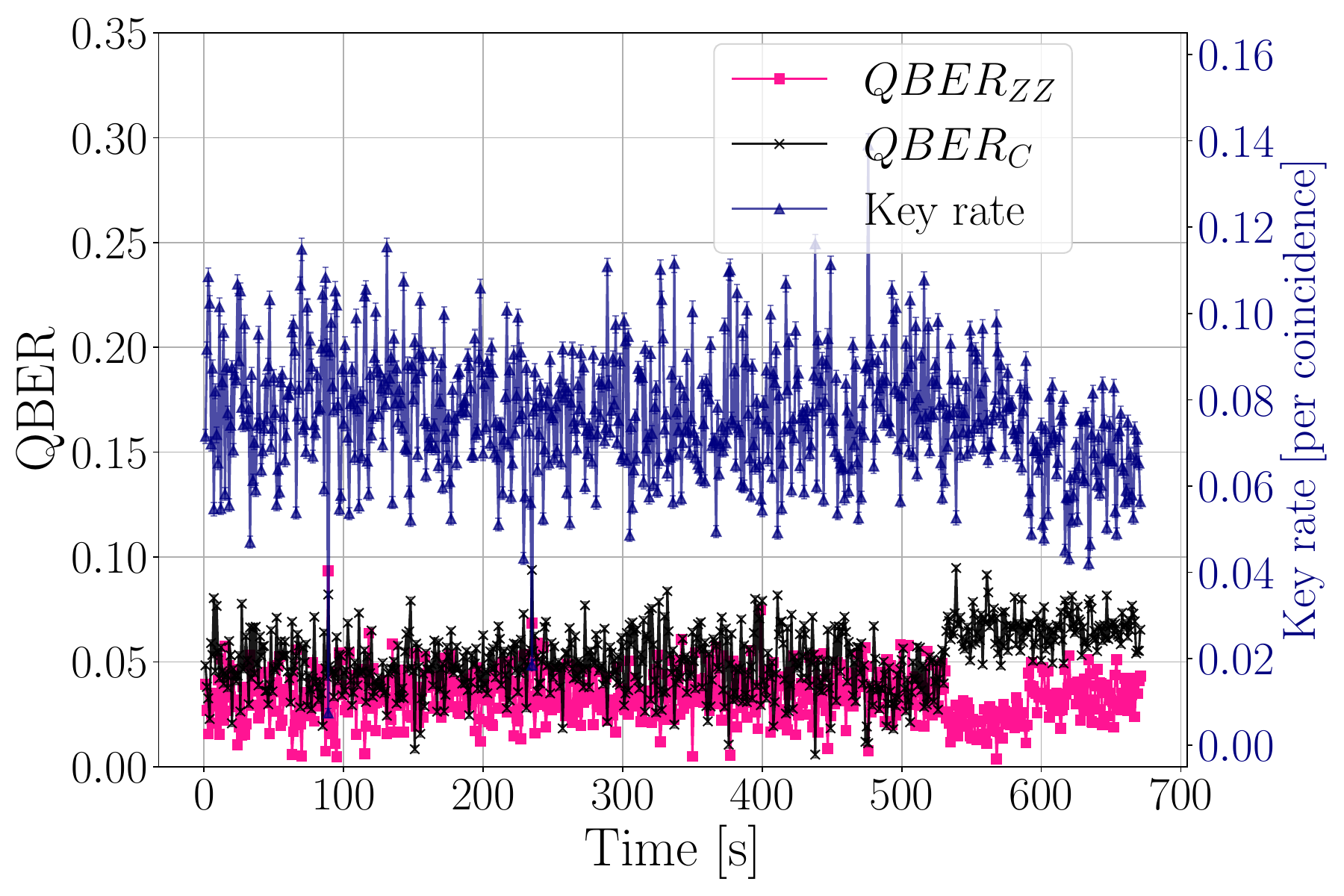}
			\caption{}
			\label{fig:key1}
		\end{subfigure}
		
		\caption{(\protect\NoHyper\subref{fig:ex1}\protect\endNoHyper) Measured correlation values and phase as a function of time for a \SI{670}{\second} link (\romannumeral 1). 
			$\phi$ is the relative phase difference between the two interferometers. (\protect\NoHyper\subref{fig:key1}\protect\endNoHyper) The corresponding asymptotic key rate and QBER. The key rate is plotted on the right-hand-side vertical axis. The integration time of each data point is \SI{1}{\second}.}
		
		\label{fig: results}
	\end{figure}
\begin{figure}[htbp]
		\centering
		\begin{subfigure}[b]{0.495\textwidth}
			\centering
			\includegraphics[width=\textwidth]{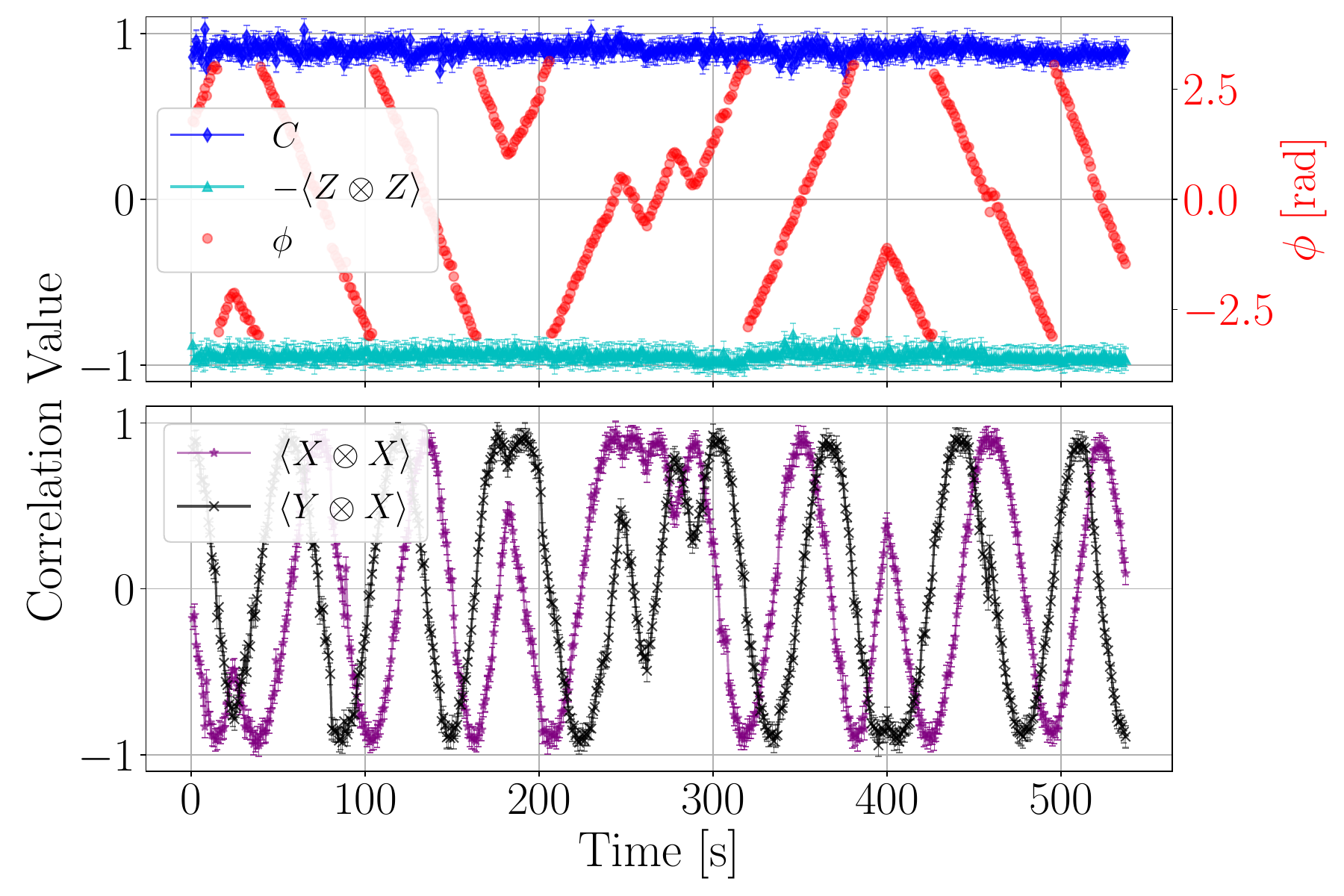}
			\caption{}
			\label{fig:ex2}
		\end{subfigure}
		\begin{subfigure}[b]{0.495\textwidth}
			\centering
			\includegraphics[width=\textwidth]{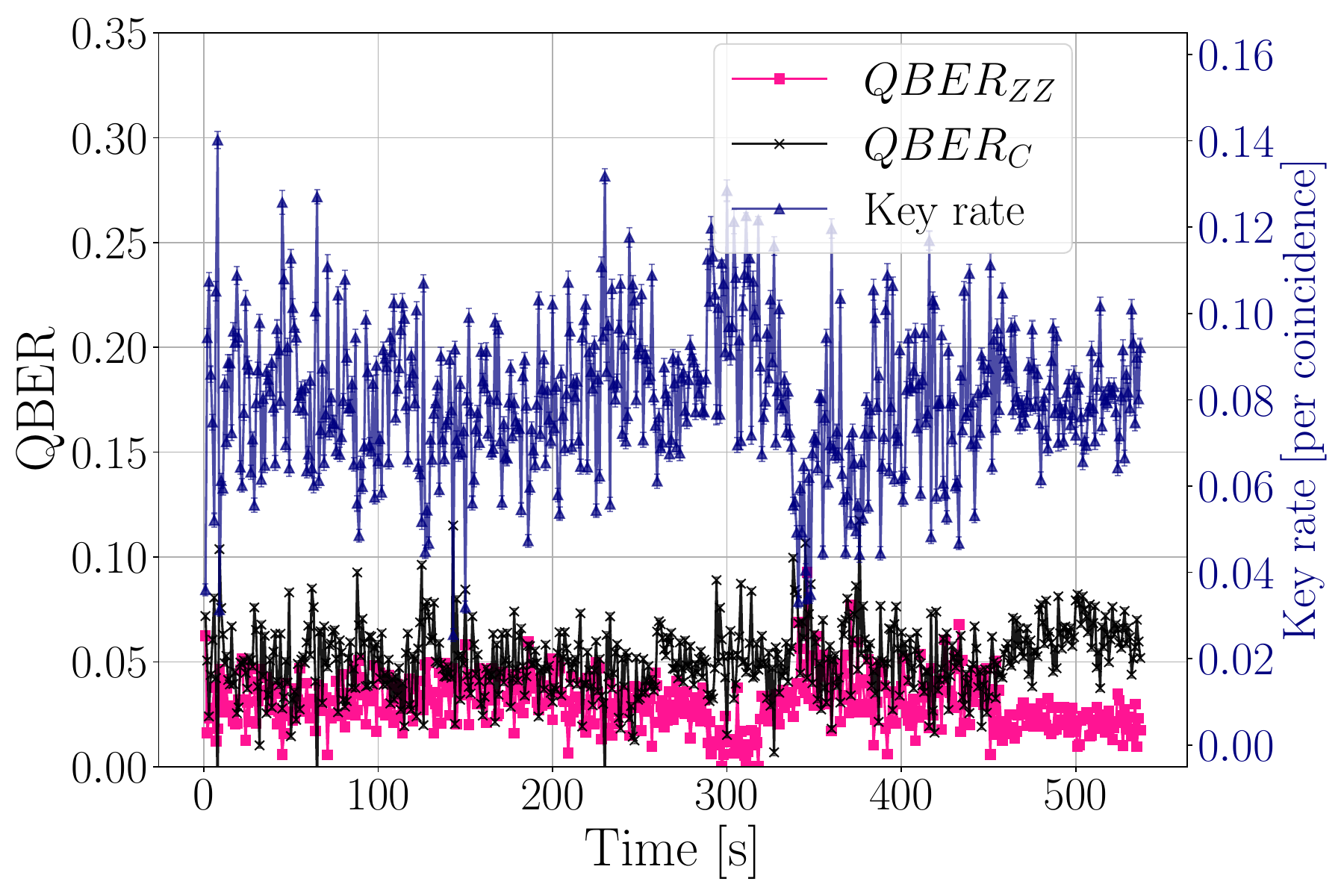}
			\caption{}
			\label{fig:key2}
		\end{subfigure}
		
		\caption{(\protect\NoHyper\subref{fig:ex2}\protect\endNoHyper) Measured correlation values and phase as a function of time for a \SI{630}{\second} demonstration while a piezoelectric actuator is inducing a phase change (\SI{0.1}{\radian\per\second}) in the TA (\romannumeral 2). 
			(\protect\NoHyper\subref{fig:key2}\protect\endNoHyper) The corresponding asymptotic key rate and QBER. The key rate is plotted on the right-hand-side vertical axis. The integration time of each data point is \SI{1}{\second}.}
		\label{fig: results2}
	\end{figure}

	We first demonstrate the feasibility of our scheme over a \SI{15}{\meter} multi-mode fiber. Here, we study two experimental scenarios; (\romannumeral 1) the system and interferometers are allowed to drift due to fluctuations from the environment, (\romannumeral 2) a piezoelectric actuator in the TA is driven to produce a constant phase change of \SI{0.1}{\radian\per\second}. The piezoelectric actuator is driven with sufficient voltage to cover a phase range of $>2\pi$, as observed from the dynamics of the expectation values in Fig.~\ref{fig: results2}(\subref{fig:ex2}). The results of the two demonstrations are shown in Fig.~\ref{fig: results} and Fig.~\ref{fig: results2}, respectively. Each data point in Fig.~\ref{fig: results},~\ref{fig: results2} is calculated with the accumulation of \SI{1}{\second} of experimental data. Coincidences are temporally filtered using a \SI{1}{\nano\second} window for each time-bin.
	Average correlation values of $C_{\romannumeral 1}=0.898\pm0.003$ and $C_{\romannumeral 2}=0.898\pm0.003$ are observed for scenarios (\romannumeral 1) and (\romannumeral 2) respectively. This corresponds to an average QBER in the superposition basis of $5.11\pm0.08\%$ and $5.09\pm0.07\%$ for (\romannumeral 1) and (\romannumeral 2) respectively, while the QBER in the computational basis is $3.34\pm0.09\%$ and $3.02\pm0.11\%$. 
	
	The visibility of the superposition basis is derived from the $C$-parameter, which is a combination of the visibility of the EPS ($95\%$), the PTC interferometer ($99\%$), and the TA interferometer ($96\%$). Combined, the estimated visibility for the superposition basis is roughly $90\%$ and gives a combined expected QBER of approximately $5\%$, which agrees with the experimentally observed values. Other errors, such as waveplate errors and imperfections in the beam splitters are not accounted for in this estimate but can contribute to the error in our demonstration. Furthermore, the estimated visibility agrees well with experiments performed without a multi-mode channel, i.e. directly connecting the output of PTC to the input TA with a single mode polarization maintaining fiber. A short \SI{1}{\meter} multi-mode fiber is still used for the input of the TA, thus in Table~\ref{tab:res} these are referred as \SI{1}{\meter} MMF experiments. On average, the visibilities of these experiments in the superposition basis of around $90\%$ and had comparable QBER and key rate values to the \SI{15}{\meter} MMF channel experiments, see Table~\ref{tab:res}.

	With the sufficiently high $C$-parameter, an asymptotic key rate was calculated following the analysis of Ref.~\cite{tannous2019demonstration}. The number of secret bits per coincidence detection at the asymptotic limit is estimated to be \SI{0.076\pm0.002}{bits\per\text{coincidence}} for (\romannumeral 1) and \SI{0.080\pm0.002}{bits\per\text{coincidence}} for (\romannumeral 2), compared to the maximum achievable rate for our system of \SI{0.167}{bits\per\text{coincidence}}. However, despite the varying phase, there is no significant variation in QBER and asymptotic key rate in Fig.~\ref{fig: results}(\subref{fig:key1}) ,~\ref{fig: results2}(\subref{fig:key2}) that is directly attributed to the phase fluctuations. In fact, the $C$-parameter is shown to be phase independent and constant throughout the entire photon exchange. This is achieved despite the deliberate phase variation between the PTC and TA (Fig.~\ref{fig: results2}(\subref{fig:ex2})).
	
	
\begin{figure}[htbp]
		\centering
		\begin{subfigure}[b]{0.495\textwidth}
			\centering
			\includegraphics[width=\textwidth]{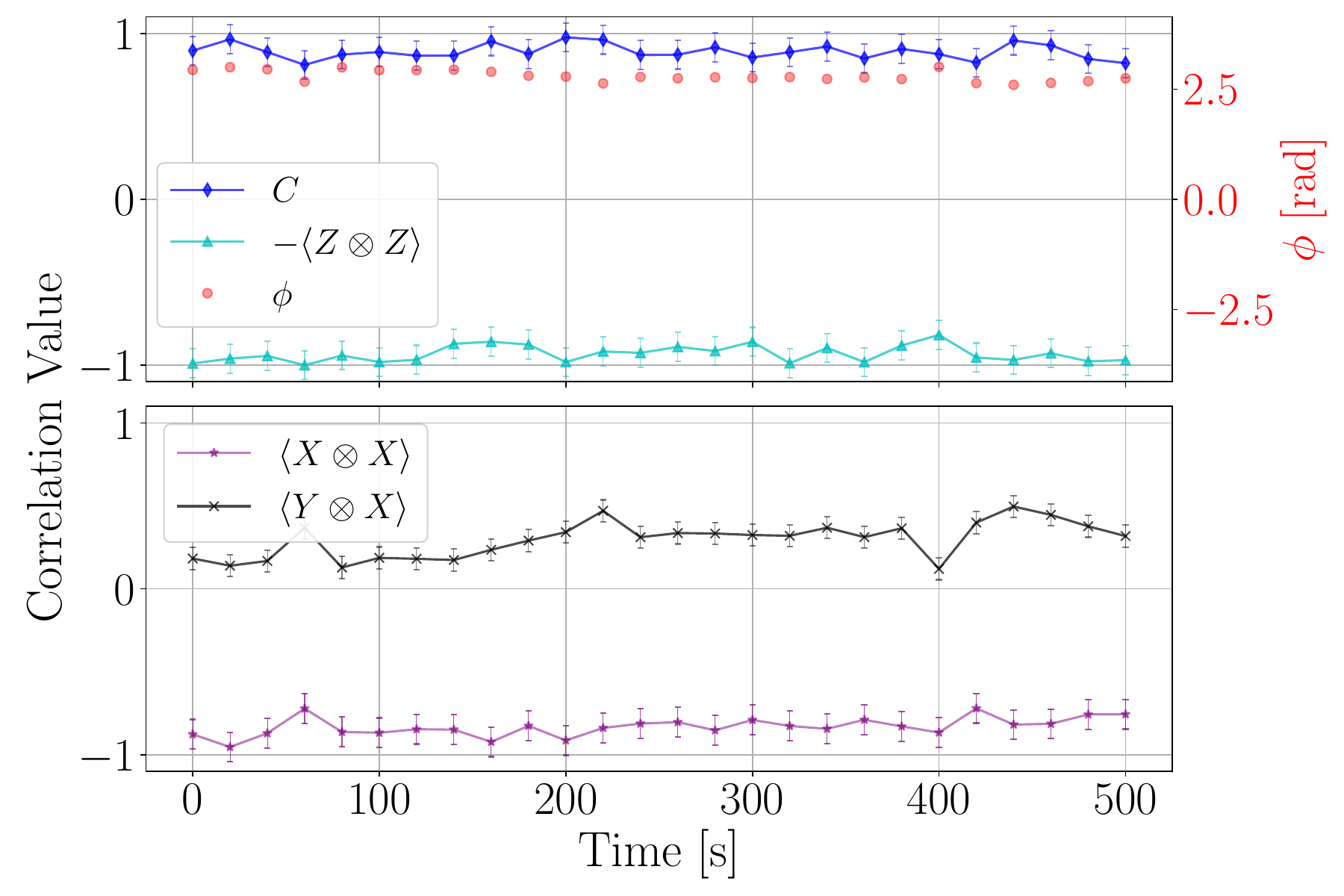}
			\caption{}
			\label{fig:fs_ex1}
		\end{subfigure}
		\begin{subfigure}[b]{0.495\textwidth}
			\centering
			\includegraphics[width=\textwidth]{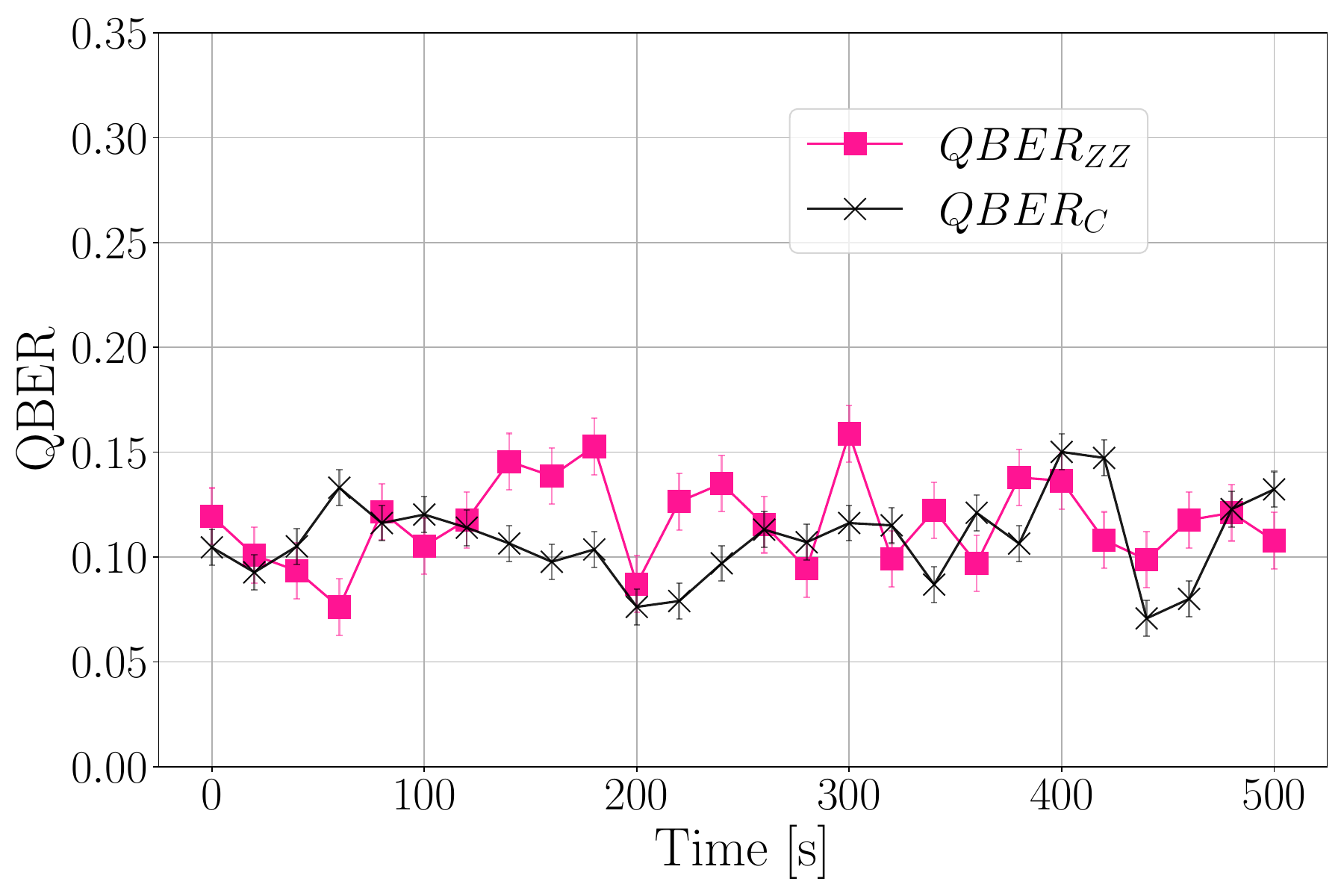} 
			\caption{}
			\label{fig:fs_qber1}
		\end{subfigure}
		\begin{subfigure}[b]{0.495\textwidth}
			\centering
			\includegraphics[width=\textwidth]{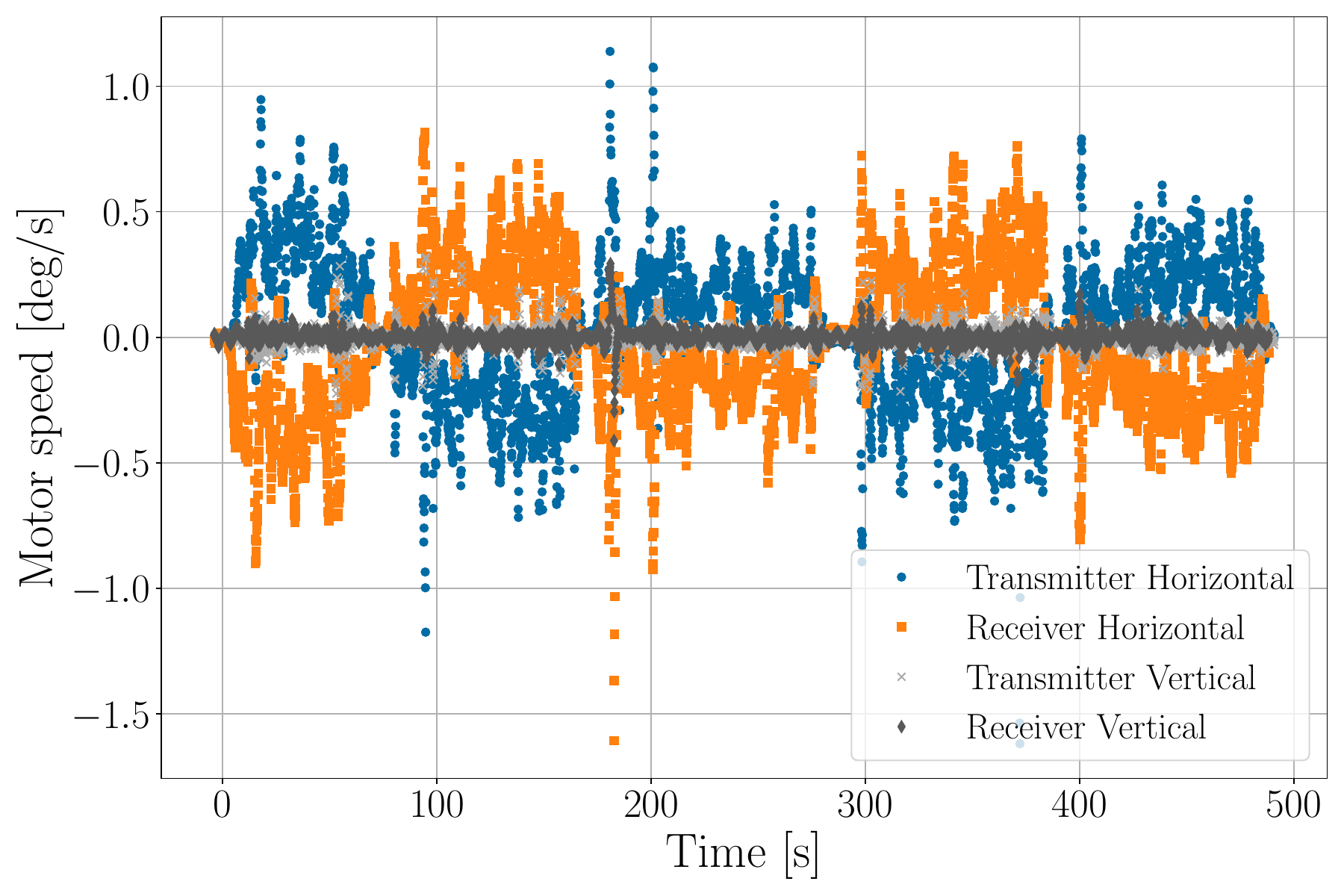}
			\caption{}
			\label{fig:fs_moving}
		\end{subfigure}
		\begin{subfigure}[b]{0.495\textwidth}
			\centering
			\includegraphics[width=\textwidth]{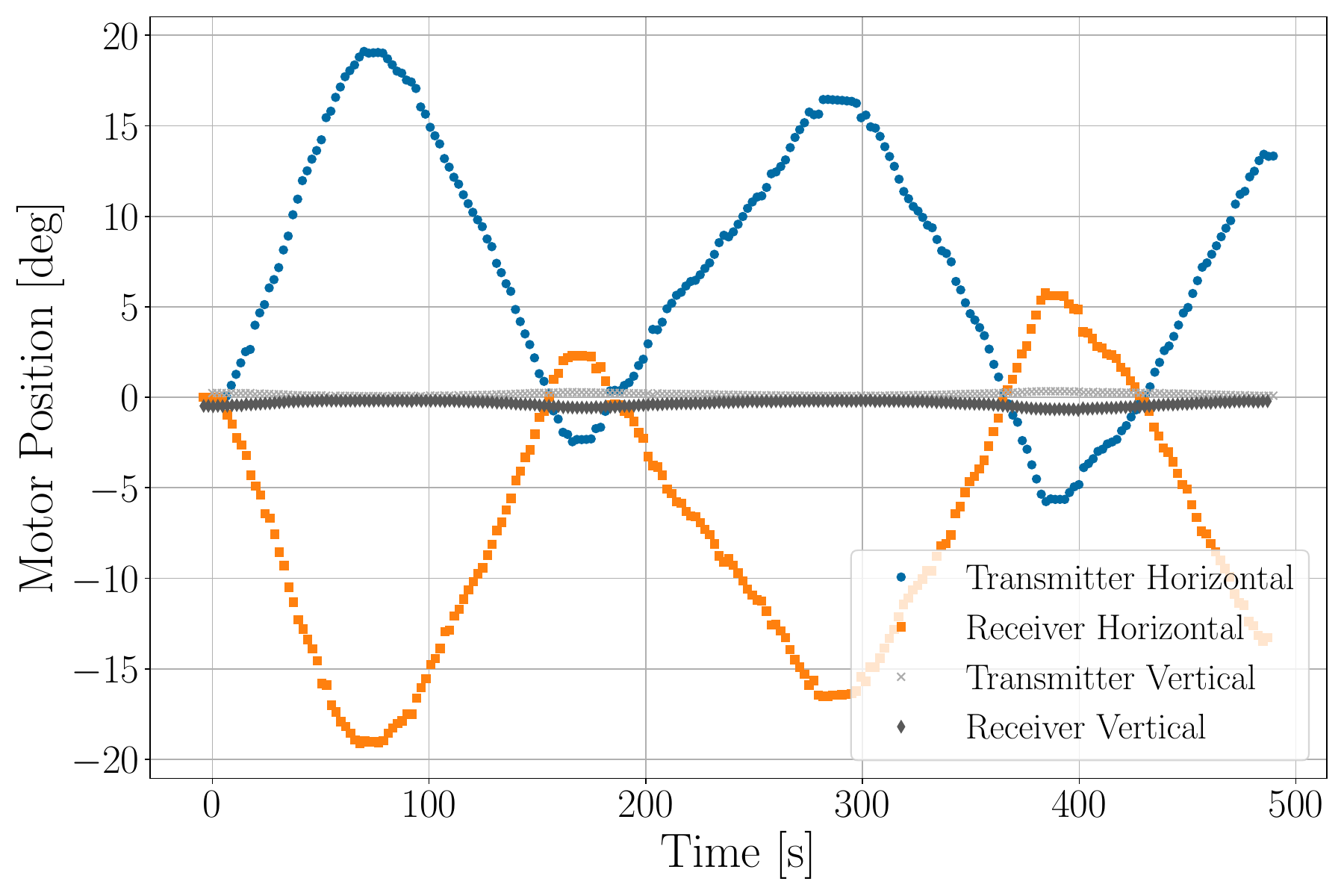}
			\caption{}
			\label{fig:fs_moving_pos}
		\end{subfigure}
		
		\caption{(\protect\NoHyper\subref{fig:fs_ex1}\protect\endNoHyper) Measured correlation values and phase as a function of time for the moving free-space channel. $\phi$ is the relative phase difference between the two interferometers. (\protect\NoHyper\subref{fig:fs_qber1}\protect\endNoHyper) The corresponding QBER. (\protect\NoHyper\subref{fig:fs_moving}\protect\endNoHyper) Applied motor speed for the horizontal and vertical axis. (\protect\NoHyper\subref{fig:fs_moving_pos}\protect\endNoHyper) Motor position throughout the experiment. Every $20^{\text{th}}$ data point is shown.}
		\label{fig:fs_results}
	\end{figure}

	With the success of our scheme over the MMF channel, we now turn to the results of the high loss moving free-space experiments (Fig.~\ref{fig:fs_results}). The system pointing data of the transmitter and receiver is shown in Fig.~\ref{fig:fs_results}(\subref{fig:fs_moving}) and~\ref{fig:fs_results}(\subref{fig:fs_moving_pos}). Each data point in Fig.~\ref{fig:fs_results}(\subref{fig:fs_ex1}),~\ref{fig:fs_results}(\subref{fig:fs_qber1}) represents the accumulation of \SI{20}{\second} of experimental data, while Fig.~\ref{fig:fs_results}(\subref{fig:fs_moving}) is recorded every \SI{0.1}{\second}. Fig.~\ref{fig:fs_results}(\subref{fig:fs_moving_pos}) is recorded at a rate of \SI{0.1}{\second} but only every $20^{\text{th}}$ point is shown for clarity. The maximum angular speed of the transmitter and receiver telescopes is \SI{1.61}{\degree\per\second} with an average angular speed of \SI{0.22}{\degree\per\second}. These numbers are comparable to expected angular velocities when tracking a Low Earth Orbit satellite. The changes in direction in the horizontal motor speed in Fig.~\ref{fig:fs_results}(\subref{fig:fs_moving}) corresponds to the change in direction of travel (Fig.~\ref{fig:fs_results}(\subref{fig:fs_moving_pos})) of the receiver when reaching the end of the rail. These moving tests are not capable of attaining the velocities required to have a measurable Doppler phase. However, they provide a proof-of-concept demonstration of a time-bin moving platform that uses MMF as the input to the measurement interferometer.

	
\begin{figure}[!htbp]
		\centering
		\includegraphics[width=0.65\columnwidth]{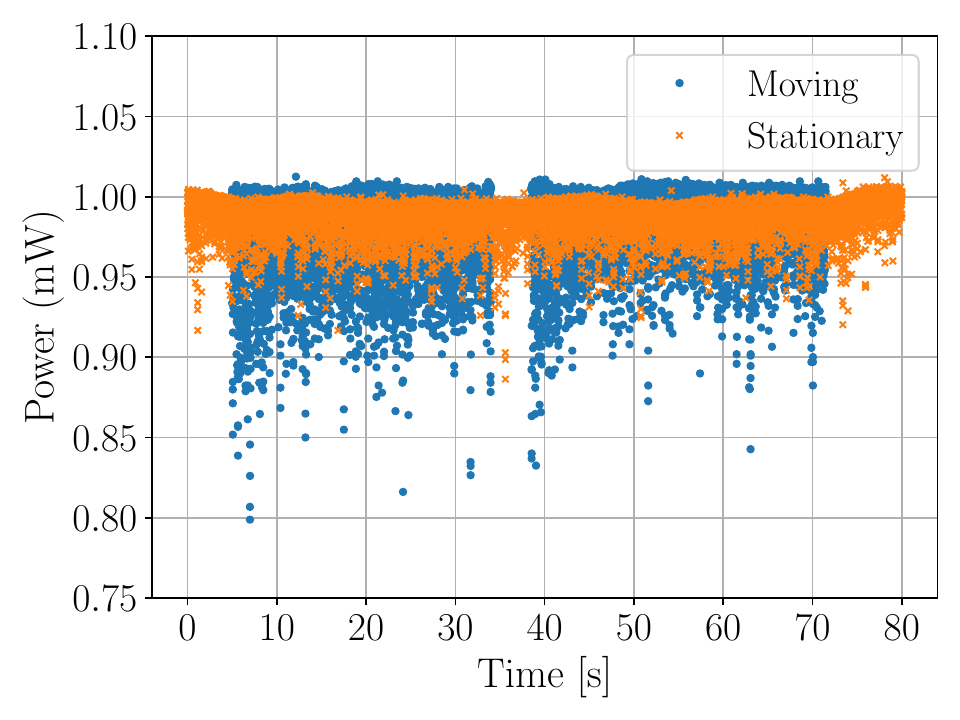}
		\caption{Stability of the TA while the receiver is moving along the track and while stationary. Note that the moving data is split into to sections, from \SIrange{5}{35}{\second}, and \SIrange{38}{72}{\second}, this corresponds each direction of motion along the rail. The intensity is measured using a Thorlabs PM100D.}
		\label{fig:stat_vs_move}
	\end{figure}

	Although the tracking system maintained signal throughout the experiment, the very low signal to noise ratio caused a high error rate of roughly $11\%$ in each basis (Fig.~\ref{fig:fs_results}(\subref{fig:fs_qber1})). Therefore no asymptotic key was calculated for the moving free-space channel. Nonetheless the correlations are measured to be $C=0.78\pm0.02$ in the phase-dependent superposition basis, and $\braket{Z\otimes Z}=0.767\pm0.027$ in the computational basis and remained relatively constant over the \SI{500}{\second} period. Note that the relative phase difference between the sender and receiver interferometer in Fig.~\ref{fig:fs_results}(\subref{fig:fs_ex1}) appears relatively stable throughout the moving tests. However, it is possible that the relative phase is moving faster than the photon detection rate permits resolving. Fig.~\ref{fig:stat_vs_move} shows the interference intensity of a long coherence laser at the output of the TA while the receiver is in motion, demonstrating how the relative phase between the two paths of the TA is changing during a round trip of the moving channel rail. Note that in Fig.~\ref{fig:stat_vs_move}, the intensity has a higher variance of \SI{0.023}{\milli\watt} while the receiver is moving compared to \SI{0.0098}{\milli\watt} for the stationary case. Despite the interferometer phase fluctuations, the visibility of the correlations in the phase-dependent basis ($C$) for the moving receiver is comparable to a stationary test over the same high loss free-space channel, $C=0.79\pm0.03$ (Table~\ref{tab:res}). Thus, the random walk of the phase drift over the \SI{20}{\second} integration time in the moving experiments is likely not the major contributor to the reduced $C$. The reduction in $C$ is dominated by the losses that reduce the signal-to-noise ratio and consequently increase the QBER.
	
	To further demonstrate that the interferometer phase fluctuations do not contribute to QBER, we subtract the background noise from our measurements and obtain very high correlations in both the computational and phase dependent superposition basis. The background is measured by offsetting the coincidence window to a time where real coincidence counts are not expected, this background is then removed in the expected coincidence window. We stress that the background subtraction is for demonstration purposes and not used for secret key generation. Nonetheless, despite the moving receiver, the multi-mode fibers, and the lack of phase stabilization we achieve correlation values of $\braket{Z\otimes Z}=0.933\pm0.030$ and $C=0.890\pm0.019$. The QBER values are significantly reduced in both bases, $3.37\pm0.75\%$ and $5.50\pm0.48\%$ respectfully. A summary of the experimental results can be found in Table~\ref{tab:res}.

	
	\begin{table}[]
		\centering
		\caption{Summary of experimental results.  FS: free-space link. B: background subtraction performed. PM: linear phase modulation is applied. A-KR: asymptotic key rate.}
		\label{tab:res}
		\begin{tabulary}{\textwidth}{L|C|C|C|C|C}
			\toprule
			Link type    & C     & QBER$_{C}$ [\%]     & $\braket{Z\otimes Z}$ & QBER$_{ZZ}$ [\%]  & A-KR [bit/coin]        \\ \midrule
			15m MMF   & 0.898(3)    & 5.11(7) & 0.933(4) & 3.3(1) & 0.076(2) \\
			15m MMF-PM   & 0.898(3)    & 5.09(7) & 0.940(4) & 3.0(1) & 0.080(2) \\ \hline
			1m MMF & 0.900(2) & 5.00(6)  & 0.937(4) & 3.1(1) & 0.079(2)  \\
			1m MMf-PM & 0.893(3)    & 5.35(8) & 0.941(7) & 2.9(2) & 0.079(3) \\ \hline
			FS   & 0.79(3)    & 10(1) & 0.84(5) & 8(1) & 0.004(16) \\
			FS moving& 0.78(2)    & 10.8(4) & 0.77(3) & 11.7(7) & - \\ \hline
			FS-B   & 0.88(4)    & 5(1) & 0.96(6) & 2(1) & 0.09(3) \\
			FS-B moving & 0.89(2)    & 5.5(5) & 0.93(3) & 3.4(7) & 0.08(1) \\ 
			\bottomrule
		\end{tabulary}
	\end{table}

	%
	To investigate the tolerances of our scheme to the effects of a rapidly changing $\phi_{AB}$, we use a Monte Carlo simulation. We consider $N$ signals sent by the EPS to Alice and Bob, and assume they arrive evenly spaced over a time period $t_N$. The instantaneous phase of the entangled state is $\phi(\tau n)$, where $n \in [0,N]$, and $\tau=t_N/N$. Numerically using the Qutip package~\cite{johansson2012qutip,JOHANSSON20131234}, we produce $N$ entangled states (\ref{eq:Estate}) each with a phase $\phi(\tau n)$. Then for each $n$ entangled state, we calculate the detection probabilities of the positive operator-valued measure (e.g. $XX_{++}$, $XX_{+-}$, $XX_{-+}$, $XX_{--}$) for each basis that is involved in calculating the $C$-parameter. Since $\phi(t)$ is changing, these probabilities should vary with time (and phase). With the probabilities calculated from every $n^\text{th}$ entangled state, a pseudorandom number generator determines the detection results of a single photon event. This is done independently for each event. The result is then counted and statistics are accumulated for the total $N$ entangled pairs. Once complete, an experimental expectation value approximation is calculated using equation~(2) of~\cite{tannous2019demonstration} and is done for each basis from which a value for the average $C$-parameter over the time interval $t_N$ is determined. This process is repeated $i$ times for each $\phi(t_N)$ and the average of the $i$ trials is taken to be the value of $C(t_N)$, with error bounds being the standard deviation of the mean. 
	
\begin{figure}[!h]
		\centering
		\begin{subfigure}[b]{0.495\textwidth}
			\centering
			\includegraphics[width=\linewidth]{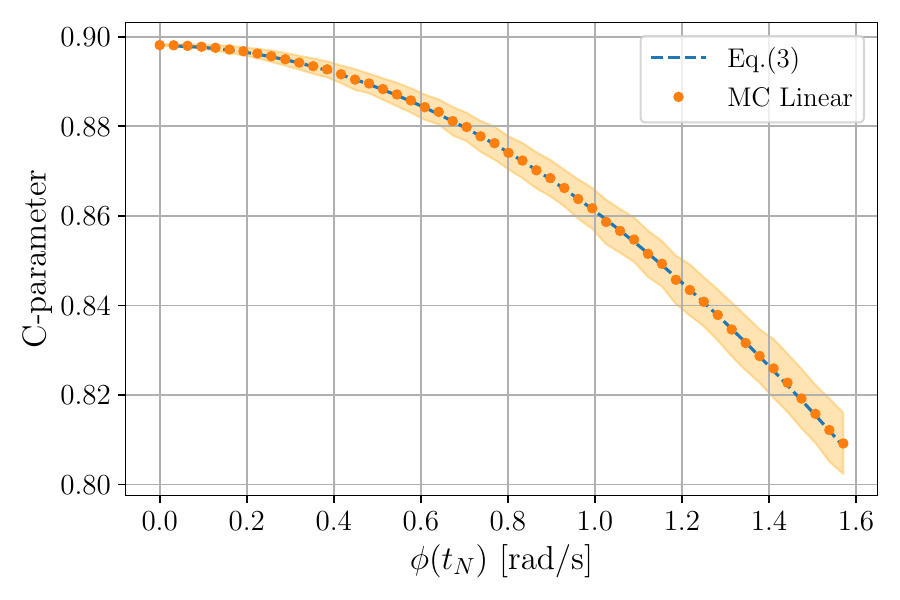}
			\caption{}
			\label{fig:linear_change}
		\end{subfigure}
		\begin{subfigure}[b]{0.495\textwidth}
			\centering
			\includegraphics[width=\linewidth]{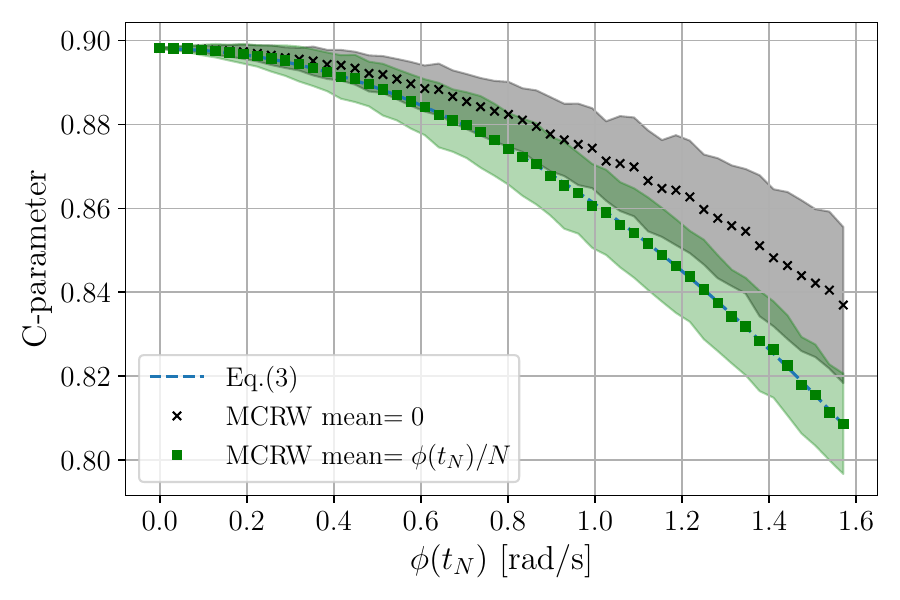}
			\caption{}
			\label{fig:rw_change}
		\end{subfigure}
		\caption{Change in the value of the $C$-parameter as a function of phase variation over a fixed time period. (\protect\NoHyper\subref{fig:linear_change}\protect\endNoHyper) A constant linear phase change as described by (\ref{eq:phaseChangeC}). (\protect\NoHyper\subref{fig:rw_change}\protect\endNoHyper) The phase change is dictated by random walks with a mean step size of zero and random walks with a mean step size of $\phi(t_N)/N$. Each walk has a standard deviation of $\phi(t_N)/\sqrt{N}$. The shaded region is the standard deviation for the distribution of the results from the Monte Carlo simulation and demonstrate the statistical nature of the $C$-parameter. Here $i=500$ for each $\phi(t_N)$ and $\alpha=170$ to match the experimental maximum visibility. For the random walk simulations, a unique walk is calculated for each $i$ trial of the simulation. MC:~Monte Carlo, RW:~random walk. }
		\label{fig:C-moving}
		
	\end{figure}

	The results of the simulation for our experimental parameters, about \SI{3000}{coincidences} ($N=3000$) in the superposition basis during a \SI{1}{\second} ($t_N=1$) measurement time interval, are shown in Fig.~\ref{fig:C-moving}. The tolerated relative phase drift is approximately \SI{0.512}{\radian\per\second} before a $5\%$ reduction in the asymptotic key rate. Comparing this to the intrinsic phase change in the interferometers of \SI{0.6}{\milli\radian\per\second} (or \SI{2.2}{\radian\per\hour}) that was measured over a 2 hour period, it is clear that our scheme can comfortably handle the intrinsic phase change of the interferometers. In Fig.~\ref{fig:C-moving}(\subref{fig:linear_change}), our Monte Carlo simulation is shown to match a modified version of equation (17) from Ref.~\cite{sheridan2010finite},
	
	\begin{equation}
		\label{eq:phaseChangeC}
		C(t_N)=C(0)\frac{\sin{\frac{\phi_N}{2}}}{N\sin{\frac{\phi_N}{2N}}},
	\end{equation}
	where $C(0)=N/(N+2\alpha)$ is related to the signal-to-noise ratio and QBER of the system when no phase change is present (i.e. the maximum value of $C$); $N$ is the number of photon detections in the time interval $t_N$; $\alpha$ is the noise counts collected in $t_N$; and $\phi_N$ is a short form notation of $\phi(t_N)$. With (\ref{eq:phaseChangeC}) direct comparisons to other systems are achievable. The modifications are necessary to match the definition of $C$ used in the 6-state 4-state protocol~\cite{tannous2019demonstration}. A detailed derivation of (\ref{eq:phaseChangeC}) and further information on the Monte Carlo simulations can be found in the supplemental materials. 
	
	In Fig.~\ref{fig:C-moving}(\subref{fig:rw_change}), we perform a Monte Carlo simulation where $\phi(t)$ changes via random walks rather than a constant linear phase as in Fig.~\ref{fig:C-moving}(\subref{fig:linear_change}). Our simulations are done using $i$ unique random walks that have a phase step distribution with a standard deviation of $\phi(t_N)/\sqrt{N}$ and a mean step size of zero (black x's), and $i$ unique random walks that has standard deviation of $\phi(t_N)/\sqrt{N}$ and a mean step size of $\phi(t_N)/N$ (green squares). The latter is effectively a linear phase drift with Gaussian noise applied to each step. In both cases the performance of the protocol in face with random walks is similar to or better than the linear case. This is because the calculation of the $C$-parameter is averaged over the measurement period, and any random walk will, on average, have an absolute phase change that is similar to the linear case. Thus, the linear phase change can be regarded as a unique case for the random walk. 
	
	For satellite free-space links, the phase tolerance of the protocol is beneficial in overcoming the relativistic Doppler phase change that occurs during a satellite pass. For a typical low Earth orbit (\SIrange{200}{1500}{\kilo\meter} in altitude), a phase change of $30\pi$ can be induced depending on the satellite pass and time-bin separation~\cite{space_super_zeitler2016super,vallone2016interference,chapman2020time}. In Ref.~\cite{vallone2016interference}, the maximum phase change experienced during the satellite pass was on the order of \SI{1}{\radian\per\second}, which would cause a $4.5\%$ drop in the C-parameter values and a $16\%$ drop in asymptotic key rate for our experimental setup. Although this phase change can be characterized and compensated with active components, the use of our passive approach reduces the overhead required. Nonetheless, the current optical setup can compensate for the Doppler phase change using the piezoelectric actuator in the TA. However, we stress that it is possible for the entire Doppler phase shift to be tolerated by the protocol if a high quality system is used, i.e. high resolution detectors and a high coincidence count transmission relative to the phase change rate. 
	
	
	
	Finally, we perform an indicative finite key rate estimation using the approach from Ref.~\cite{sheridan2010finite} that analyzed a 6-state entanglement-based RFI protocol. Note that this is not a complete finite size analysis nor a key generating post-processing scheme (i.e. no error correction, privacy amplifications, etc.). For our analysis, we make the assumption that the correlations of the entangled state shared between Alice and Bob are symmetric ($\braket{X_A\otimes X_B}^2=\braket{Y_A\otimes Y_B}^2$ and $\braket{Y_A\otimes X_B}^2=\braket{X_A\otimes Y_B}^2$) such that we can take the 6-state 4-state $C$-parameter ($C_{64}$) and relate it to the 6-state $C_{66}$ as $C_{66}=2C^2_{64}$. Proper investigation of this assumption is a topic of further study. In addition, future work will be done to use modern numerical finite size key analysis~\cite{george2021numerical}. Using equations (2) and (3) from Ref.~\cite{sheridan2010finite}, the expression for the secret-key length secure against coherent attack for our system given by
	\begin{eqnarray}
		r_N=\frac nN[1-I_E(Q',C')-n_Qfh(Q')-\log_2\frac{2}{\varepsilon_{\text{EC}}}-2\log_2\frac{1}{\varepsilon_{\text{PA}}}-7\sqrt{\frac{\log_2\left(2/\bar{\varepsilon}\right)}{n}}\nonumber\\
		-\frac{30}{N}\log_2\left(N+1\right)],\label{eq:finite_kr}
	\end{eqnarray}
	where $N$ is the total raw key (signals sent by Alice and detected by Bob), $n$ is the number of detections in the key map (computational) basis, $C'$ ($Q'$) is the observed $C$-parameter (error rate in the key map basis) modified to account for finite statistics, and $I_E$ is the upper bound on the information that Eve can gain and is given by equation (9) of Ref.~\cite{sheridan2010finite}. The other key rate analysis parameters are found in Table~\ref{tab:finite_params}. For the multi-mode fiber demonstration (\romannumeral 1), we analyzed the results from the entire \SI{670}{\second}. This gave a total number of coincidences of $4464522$, a computational basis QBER of $3.3\%$ and a $C_{64}$-parameter of $0.8909$. With these parameters and assuming perfect error correction, we calculate a secret key fraction of $0.0546$, which amounts to $243748$ secret bits from the total raw key. For (\romannumeral 2), considering the entire \SI{630}{\second} link would not produce any key as the smearing of the $C$-parameter would be too great. However, shorter block sizes would risk not producing positive key due to low photon statistics. For an error correction factor of $f=1.2$, no positive secret key is achievable due to the low photon statistics relative to the phase change rate. A complete finite size analysis that utilizes the $C$-parameter, as in the asymptotic case, is a topic of future study. Regardless, the current coincidence detection rates would need to be improved to reach values that are comparable with higher performing systems reported in the literature~\cite{feng_four-state_2021,tang_free-running_2022}.
	
	\begin{table}[h]
		\caption{Parameters of the finite size secret key analysis.}
		\label{tab:finite_params}
		\centering
		\begin{tabulary}{0.4\textwidth}{c|l|c}
			\toprule
			Parameter & Description & Value\\
			\midrule
			$\bar{\varepsilon}$ & Smooth min entropy estimation error & $2.5\times10^{-9}$ \\
			$\varepsilon_{\text{EC}}$ & Error correction failure probability & $2.5\times10^{-9}$ \\
			$\varepsilon_{\text{PA}}$ & Privacy amplification failure probability & $2.5\times10^{-9}$ \\
			$\varepsilon_{\text{PE}}$& Parameter estimation failure probability & $2.5\times10^{-9}$ \\
			$n_{Q}$ & Fraction of computational basis used to estimate the QBER & $0.1$\\
			$f$& Error correction factor & 1.2\\ \bottomrule
		\end{tabulary}
	\end{table}

	\section{Discussion and Conclusion}
	To our knowledge, our results are of the first fully passive time-bin RFI-QKD implementation over a multi-mode fiber channel and time-bin entanglement correlations to a moving receiver. We demonstrated time-bin QKD without the need for any active phase stabilization of the sender and receiver interferometers, and without actively correcting for spatial mode distortions. While using a hybrid entanglement-based 6-state 4-state RFI-QKD protocol, we observed error rates of less than $5.5\%$ in the superposition basis, and a continuous asymptotic key rate of above \SI{0.07}{bits\per\text{coincidence}} over a \SI{15}{\meter} multi-mode fiber. Over the high loss moving-free space channel experiment we observed constant entanglement correlations while the receiver was in motion.
	
	Combining the results of both experiments, we have demonstrated a proof-of-concept implementation of employing a multi-mode fiber as the input for a time-bin decoding interferometer. This enables a significant reduction in the required need for adaptive optics for a receiver in a moving link compared to using a single-mode fiber interferometer similar to those used in previous static free-space RFI-QKD demonstrations~\cite{chen_field_2020}. The advantage of using a multi-mode fiber is particularly relevant for platforms where resources are limited due to better pointing requirements and optical transmission, especially in turbulent free-space channels~\cite{fabianFSMMF}. Our purely passive RFI scheme is particularly suitable for moving platforms where maintaining a stable phase reference can be difficult. Our approach can also passively handle some of the relativistic Doppler phase that occurs during satellite passes, further reducing the overhead of a satellite-to-ground time-bin link. 
	
	Our results are of great practical importance for the use of free-space time-bin encoded quantum channels, because the ability to distribute entanglement correlations over a highly multi-mode channels without any mode filtering, sorting, nor compensation can ease the deployment of quantum networks, entanglement distribution experiments, and quantum sensing applications~\cite{sajeed_observing_2021}. Future work is to further the theoretical studies on the robustness of the protocol, particularly against rapid phase fluctuations and finite size effects. To bring our proof-of-concept demonstration in line with other RFI-QKD implementations~\cite{liu2024experimental}, improvements will involve using a high-performance system that has a photon source that is optimized for the apparatus wavelength and bandwidth, as well as an increased EPS heralding efficiency and visibility~\cite{anwar2021entangled}. The optical losses in our demonstration need to be addressed as there is a minimum number of detections required to suitably estimate the $C$-parameter and error rates. Enhancing the brightness of the photon source and the efficiency of the free-space channel would significantly improve the signal-to-noise ratio, thereby positively impacting the other experimental parameters. Additionally, improvements to the performance of the field widened interferometer can help reduce the overall QBER, by increasing the performance and robustness of the design, by for instance using a compact optical layout~\cite{tannous2023advancing,cahall_multi-mode_2020}. Further loss reductions can be found by removing the intrinsic $50\%$ loss of the PTC by using an electronic or optical switch~\cite{england2021perspectives}. Enhancing the receiver system, particularly by removing the unnecessary losses induced by the polarization analyzer, would also be beneficial.
	
	We emphasize that while our demonstration performed over short distances, \SI{5}{\meter} moving free-space and \SI{15}{\meter} for the MMF channel, it is possible to achieve longer distances for both channels. For the free-space channel increasing the distance is straightforward. While for MMF using a low modal dispersion graded-index fiber (such as OM4) and by adjusting the time-bin separation accordingly can increase the feasible distance, but is limited by fiber losses for photons at \SI{785}{\nano\meter}. 
	Finally, while our demonstration is realized using near-infrared photons around \SI{800}{\nano\meter} which are suitable for atmospheric free-space channels~\cite{bourgoin2013comprehensive}, extending our scheme to telecommunication wavelengths, could be enabled by the recent advancement of multi-mode coupled superconducting nanowire single-photon detectors~\cite{chang2019multimode}.  

		\begin{backmatter}
			\bmsection{Funding}
			The authors are grateful for the support from the Canada Foundation for Innovation, the Ontario Research Fund, the Canadian Institute for Advanced Research, the Natural Sciences and Engineering Research Council of Canada, Industry Canada, and the Research Centres of Excellence program supported by the National Research Foundation (NRF) Singapore and the Ministry of Education, Singapore. RT would like to thank the NSERC CGS-D for personal funding.
			
			\bmsection{Acknowledgments}
			The authors would like to thank Dr. Norbert L{\"u}tkenhaus and Scott Johnstun for discussions on the analysis. 
			
			\bmsection{Disclosures}

			\medskip
			
			\noindent The authors declare no conflicts of interest.

			\bmsection{Data availability} Data underlying the results presented in this paper are not publicly available at this time but may be obtained from the authors upon reasonable request. Select data is available at \cite{DVN/3ODODO_2023}.

			\bmsection{Supplemental Materials}
			See \href{https://doi.org/10.6084/m9.figshare.29042372}{Supplement 1} for supporting content
		\end{backmatter}
			
		\bibliography{references}

		\end{document}